\documentclass[10pt,preprint2]{aastex}

\begin{document}

\title{The Cross-Wavelet Transform and Analysis of Quasiperiodic Behavior in the Pearson-Readhead VLBI Survey Sources} 
\author{Brandon C. Kelly, Philip A. Hughes, Hugh D. Aller, and Margo F. Aller}
\affil{Astronomy Department, University of Michigan, Ann Arbor, MI 48109-1090}
\email{bckelly@umich.edu, hughes,hugh,margo@astro.lsa.umich.edu}

\begin{abstract}

We introduce an algorithm for applying a cross-wavelet transform to analysis of
quasiperiodic variations in a time-series, and introduce significance tests for
the technique.  We apply a continuous wavelet transform and the cross-wavelet
algorithm to the Pearson-Readhead VLBI survey sources using data obtained from
the University of Michigan 26-m parabloid at observing frequencies of 14.5,
8.0, and 4.8 GHz.  Thirty of the sixty-two sources were chosen to have
sufficient data for analysis, having at least 100 data points for a given
time-series.  Of these thirty sources, a little more than half exhibited
evidence for quasiperiodic behavior in at least one observing frequency, with a
mean characteristic period of 2.4 yr and standard deviation of 1.3 yr.  We find
that out of the thirty sources, there were about four time scales for every ten
time series, and about half of those sources showing quasiperiodic behavior
repeated the behavior in at least one other observing frequency.

\end{abstract}


\section{INTRODUCTION}

It is well accepted that centimeter waveband emission from AGNs is associated
with a jet of synchotron plasma, the accretion structure and immediate
environment of a supermassive black hole contributing broad band emission from
the infrared to the gamma ray spectrum.  Temporal variations are observed in
the radio flux, and a number of processes have been proposed to explain this,
such as an accretion rate that may change with time, an accretion disk that
exhibits instability, an outflow that may be Kelvin-Helmholtz unstable, or an
outflow that may interact with ambient inhomogeneities.  The temporal
variations in the radio flux motivate a search for characteristic time scales,
and, if found, would lend insight into the mechanisms causing the variations.
 
In practice, searching for possible quasiperiodic behavior masked by a
stochastic component can be rather difficult, especially in the context of
irregular time sampling.  One promising technique is to perform a continuous
wavelet transform on the signal, and map out the coefficients in wavelet space.
This was done on data for the BL Lac OJ 287, resulting in evidence for periods
in the radio spectral region of $\sim$ 1.66 yr, and $\sim$ 1.12 yr dominating
in the 1980s \citep{hug98}.  The results led \citet{hug98} to propose that the
results can best be explained by a ``shock-in-jet'' model, in which the longer
periodicity is linked with an otherwise quiescent jet, and the shorter with a
passing shock-wave.
 
The continuous wavelet transform detects quasiperiodic behavior through visual
examination of the map over translation and dilation, and assigns a
characteristic time scale by looking for peaks in the time-averaged wavelet
power (see \S~4.1).  The scales that the peaks occur at correspond to Fourier
periods, from which a characteristic time scale may be deduced.  Although the
continuous transform is effective, when assigning a characteristic time scale
it does not fully include information of how a time series varies in dilation;
the time scale corresponds to those dilations where the time-averaged wavelet
power spectrum peaks. In order to use all of the available information in the
transform, we examine the cross-wavelet transform, in which the coefficients
for the continuous transform of a signal are multiplied by the complex
conjugate of the coefficients of another signal.  The results are then mapped
out in wavelet space and analyzed for correlation.  A cross-wavelet analysis
was used on the OJ-94 project light curve \citep{leh99}, and has seen
application in other areas of science as well \citep{pan00, kyp99}.  We examine
this method in greater depth, as well as a technique for finding the
characteristic period of a signal and apply this to the Pearson-Readhead (PR)
VLBI survey sources using data from the University of Michigan Radio Astronomy
Observatory (UMRAO) observed at three frequencies.  The PR group is well suited
for analyzing quasiperiodicity, due to the high signal-to-noise in the data for
most of these sources \citep{all02}, as well as investigating quasiperiodic
variations (QPVs) as a function of optical source type, since it contains
representatives from optical classes QSO, galaxy, and BL Lac.  In addition, the
time base of the UMRAO data is around twenty years for most sources, allowing
possible quasiperiodic events of as long as four years to be observed for at
least four to five cycles.  Using UMRAO data, evidence has been found for
periodic behavior in the centimeter waveband emission for the BL Lac object
0235+164 \citep{roy00}.  However, there have not been any convincing
indications of periodicity for PR sources \citep{all02}.

\section{THE DATA}

The data for this study were acquired as part of the UMRAO program, using the
University of Michigan 26-meter parabloid.  Sixty-two of the sixty-five PR
sources are within the declination limits of the parabloid, and these sources
have been observed at least trimonthly at 14.5, 8.0, and 4.8 GHz since the fall
of 1984, with occasional gaps due to poor weather and positions too near to the
sun to be observed.  The observing technique and reduction procedures used are
those described in \citet{all85} and \citet{all01}, with the latter including a
rescaling of the data for the purpose of conforming with the flux system of
\citet{ott94}.

\section{WAVELET TRANSFORMS}

\subsection{The Continuous Wavelet Transform}

The continuous wavelet transform involves decomposing a signal $f(t)$ into a
number of translated and dilated wavelets.  The main idea behind this is to
take a ``mother'' wavelet, translate and dilate it, convolve it with the
function of interest, and map out the coefficients in ``wavelet space'',
spanned by translation and dilation.  Periodic behavior then shows up as a
pattern spanning all translations at a given dilation, and this redundancy in
the wavelet space makes detection of periodic behavior rather easy.  The
wavelet transform preserves temporal locality, which is an advantage over
Fourier analysis.  For instance, power associated with irregular sampling does
not contribute to the coefficients as in Fourier analysis, which is extremely
helpful when using poorly-sampled data.

There are several common types of mother wavelets.  In this analysis we use a Morlet wavelet of $k_{\psi}=6$, given by
\begin{equation}
\psi_{\rm Morlet}= \pi^{-1/4} e^{-ik_{\psi}t} e^{-|t^2|/2},
\label{eq01}
\end{equation} 
with coefficients
\begin{equation}
\tilde f\left(l,t'\right)=\int_{ {\bf R}} f\left(t\right) \psi_{lt'}^*
\left(t\right)dt
\label{eq02}
\end{equation}
and
\begin{equation}
\psi_{lt'}\left(t\right)=\frac{1}{\sqrt{l}} \psi\left({t-t'\over l}\right), \ \  l\in
 {\bf R}^+,\ \ t\in {\bf R}.
\label{eq03}
\end{equation}
The continuous transform must also satisfy an admissibility condition, requiring zero mean:
\begin{equation}
\int_{ {\bf R}} \psi (t) dt = 0.
\label{eq04}
\end{equation} 
It should be noted that $l$ corresponds to dilations, and $t'$ refers
to translations.  Generally the parameter $k_{\psi}$ remains fixed throughout the analysis, and for simplicity will hereafter be just $k$.  Figure \ref{fig1}a shows the continuous wavelet transform for a
sinusoidal signal.  The periodic behavior is easily revealed by the redundancy
in the plot.  For the interested reader, more detailed information on the
continuous wavelet transform may be found in \citet{far92}.

\subsection{The Cross-Wavelet Transform}

The continuous wavelet transform is effective for examining how a time series varies in time and scale, but fails to include how it varies over a range of scales when assigning a period that best characterizes it.  After identifying that a periodic pattern exists in what can be potentially noisy and poorly sampled data, one finds the dilation which characterizes the period from the time-averaged data (i.e., the wavelet power spectrum), and from that dilation the period is calculated.  For a quasiperiodic signal there is no unique dilation; there is a need for a method of directly measuring a characteristic time scale, or scales, that includes information on how a source varies in dilation, and for this we shall examine the cross-wavelet transform.
 
Given two time series $f_{a} (t)$ and $f_{m} (t)$, we can construct the cross-wavelet
transform
\begin{equation}
\tilde f_{c}\left(l,t'\right)=\tilde f_{a}\left(l,t'\right) \cdot \tilde f_{m}^*\left(l,t'\right),
\label{eq05}
\end{equation}
where $\tilde f_{a}$ and $\tilde f_{m}$ are given by Equation (\ref{eq02}).  In this analysis $f_{a} (t)$ is the source signal and $f_{m} (t)$ is a sinusoidal mock signal, although one can use any two time series thought to be correlated.
 
Because we are looking for quasiperiodic behavior, first assume an ideal
signal of the form
\begin{equation}
f_{a}\left(t\right)=A_a e^{i(\omega_{a}t+\phi_a)}, \ \ \  -\infty<t<\infty,
\label{eq06}
\end{equation}
where $\omega_a=2 \pi / \tau_a$ for period $\tau_a$.
Anticipating that it will be productive to cross the actual signal with a signal of a similar
form, we try a mock signal
\begin{equation}
f_{m}\left(t\right)=A_m e^{i(\omega_{m}t+ \phi_m)}, \ \ \  -\infty<t<\infty.
\label{eq07}
\end{equation}
Each of these signals will have continuous wavelet coefficients given by Equation (\ref{eq02}):
\begin{equation}
\tilde f\left(l,t'\right)=A \sqrt{2 l \pi^{1/2}} e^{-\frac{1}{2}(l \omega - k)^2} e^{i (\omega t'+\phi)} .
\label{eq08}
\end{equation}
The cross-wavelet (Eq.[\ref{eq02}]) for these two signals becomes
\begin{eqnarray}
\tilde f_{c}\left(l,t'\right) = 2 \sqrt{\pi} A_a A_m e^{i(\phi_a-\phi_m)} l e^{-k^2} \times & & \nonumber \\
\ \ \ \ \ \ \ e^{-\frac{1}{2}[(\omega_{a}^2+\omega_{m}^2) l^2 - 2 k l (\omega_{a}+\omega_{m})]} e^{(\omega_{a}-\omega_{m})it'}. & & 
\label{eq09}
\end{eqnarray}
Defining
\begin{eqnarray*}
A & \equiv & 2 \sqrt{\pi} A_{a} A_{m} e^{-k^2} \\
\eta & \equiv & \omega_{a}^2 + \omega_{m}^2 \\
\gamma & \equiv & \omega_{a} + \omega_{m} \\
\beta & \equiv & \omega_{a} - \omega_{m}\\
\phi & \equiv & \phi_a - \phi_m,
\end{eqnarray*}
Equation (\ref{eq09}) then becomes
\begin{equation}
\tilde f_{c}\left(l,t'\right)= A l e^{-\frac{1}{2}(\eta l^2 - 2 k \gamma l)}
e^{i( \beta t' + \phi)},
\label{eq10}
\end{equation}
which is easily interpreted.  From this equation we note two important
properties of this cross-wavelet---first, it has the form of a Gaussian in
the dilation coordinate $l$, and second, it is sinusoidal in the
translation coordinate $t'$ with frequency given by the difference in
the frequencies of the actual and the mock signal (i.e., the beat frequency).  When these two
frequencies are equal, the translation dependence is lost, and the
cross-wavelet reduces to a Gaussian in the dilation coordinate.  Figure \ref{fig1}b, shows the cross-wavelet for two sinusoids oscillating with the nearly same frequency.

\subsection{Using the Cross-Wavelet Transform to Analyze Periodicity}

Motivated by the results from the previous section, we can develop an
algorithm to find the period which best characterizes a signal, i.e., the characteristic time scale.  Because we are dealing with
real signals, noise will be present, and we explore the role of noise in \S~4 and \S~5.  In addition, the values of $t$ and $t'$ do not extend
to infinity, but over the range $t=a$ to $t=b$.
Because of the finiteness of the range of $t$, the range of dilations
becomes $l_{min}<l<l_{max}$.  We can still use Equation (\ref{eq08}) as an
approximation when our range of $l$ and $t'$ is such that the wavelets  $\psi_{lt'}\left(t\right)$ contribute negligibly outside of $a<t<b$, and
\begin{equation}
\int_{a}^{b} e^{-[(\omega-\frac{k}{l})it \pm \frac{1}{2}|{t-t'\over l}|^2]}dt
\approx \int_{-\infty}^{\infty} e^{-[(\omega-\frac{k}{l})it \pm \frac{1}{2}|{t-t'\over l}|^2]}dt,
\label{eq11}
\end{equation}
 
Returning to Equation (\ref{eq10}), we define the modulus as
\begin{equation}
\left|\tilde f_{c}\left(l,t'\right)\right|^2=A^2 l^2 e^{-(\eta l^2 - 2 k \gamma l)}.
\label{eq12}
\end{equation}
Integrating over the region of
wavelet space we mapped out, weighting by the scale, and approximating the integral over $l$ out to infinity, we arrive at
\begin{eqnarray}
F_c(\tau_m) & = & \int_{l_{min}}^{l_{max}} \int_{a}^{b} \frac{\left| \tilde f_{c}\left(l,t'\right)
\right|^2}{l} dt' dl \nonumber \\
 & \approx & A^2 (b-a) \frac{k \gamma}{\eta} \sqrt{\frac{\pi}{\eta}} e^{{k^2 \gamma^2
\over \eta}}.
\label{eq13}
\end{eqnarray}
We have empirically found that this function attains it maximum value at $\tau_a=0.973 \tau_m$ for a value of $k=6$.  This result is not surprising, as one would expect the power of the cross coefficients to reach their maximum when the periods for the two signals are almost equal.  One can think of the mock signal as acting like a filter in wavelet space; however, the power of the mock coefficients, $|\tilde f_m|^2$, are slightly asymmetric about the peak (see Eq.[\ref{eq08}]).  The weighted total power of the cross coefficients, $F_c$, does not attain its maximum exactly at $\tau_m=\tau_a$ because of this slight asymmetry.

To avoid confusion, one must remember that previously when the integration was over $[a,b]$ we integrated over the original coordinate $t$ and, because the wavelets have compact support in $t$, we approximated the integral by taking the limits to be $[-\infty,\infty]$.  Here, however, the integral is over the translation coordinate $t'$, so we must restrict the limits to $[a,b]$.  In addition, 
the approximation of the integral over $l$ is valid so long as $\tilde f_{c}(l,t') / l$ does not peak in the dilation coordinate too close to $l_{min}$ or $l_{max}$.  For our purposes, $l_{min}=2 \delta t$ and $l_{max}=N_{data} \delta t /3$, where $N_{data}$ is the number of data points and $\delta t$ is the time spacing.  Typical values of $b-a$ for the PR Survey Sources are roughly 20 years, and we require at least $N_{data}=100$ data points for a reliable cross-wavelet analysis, which results in $0 < l_{min} \leq 0.4$ yr and $l_{max} < 12$ yr.  We search for periods within $0.5< \tau < (b-a)/4$ years.  We have inspected $|\tilde f_c|^2 / l$ graphically to assure that the approximation is valid for periods $\tau_a \ge 0.7$ yr for all sources having more than 100 data points.

In practice, one must worry about edge effects as the time series is finite.  Because the convolution of Equation (\ref{eq02}) introduces power into the wavelet coefficients $\tilde f$ from the discontinuity at the edges of the time series, there is a region where the wavelet coefficients will be contaminated from edge effects.  This region is known as the {\it cone of influence}.  The wavelet power associated with edge effects becomes negligible for translations $t'$ farther than $\sqrt{2} l$ from the edge \citep{tor98}

One more useful quantity is the point $\tilde l$ where $\tilde f_{c}(l,t')$
peaks in the dilation coordinate.  This is given by
\begin{equation}
\tilde l={k\left(\omega_{a}+\omega_{m}\right)+\sqrt{(k^2+4)(\omega_{a}^2
+\omega_{m}^2)+2 k^2 \omega_{a} \omega_{m}} \over 2(\omega_{a}^2+\omega_{m}^2)}.
\label{eq14}
\end{equation}
When $\omega_{m}=\omega_{a}$, this becomes
\begin{equation}
\tilde l={k + \sqrt{k^2 + 2} \over 2 \omega_{a}},
\label{eq15}
\end{equation}
or,
\begin{equation}
\tau_a = \frac{4 \pi \tilde l}{k + \sqrt{k^2 + 2}} , \ \ \tau_a = \frac{2 \pi}{\omega_a}.
\label{eq16}
\end{equation}
This equation shows the relationship between a scale $l$ and the corresponding Fourier period, and can also be arrived at by inputing a sinusoid for $f(t)$ in Equation (\ref{eq02}).  For clarification, we will use the term `Fourier period' when referring to the period associated with a certain scale $l$, and the terms `analyzing period' or `mock period' when referring to the period $\tau_m$ associated with the mock signal.

\section{SIGNIFICANCE TESTS}

\subsection{Tests for the Continuous Transform}

Wavelet analysis has historically suffered from a lack of statistical significance tests.  In \citet{tor98} an excellent discussion of statistical significance for the continuous wavelet transform is given and supported by Monte Carlo results, and we summarize the relevant points.  When implementing the wavelet transform, one must use sums and discrete points rather than the theoretical treatment given earlier, and we switch to using these.

The model of correlated noise most likely to closely resemble the UMRAO data is the univariate lag-1 autoregressive (AR(1)) process \citep{hug92}, given by
\begin{equation}
x_n = \alpha x_{n-1} + z_n,
\label{eq17}
\end{equation}
where $\alpha$ is the assumed lag-1 autocorrelation and $z_n$ is a random deviate taken from white noise (note that this is a `first order' process).  The normalized discrete Fourier power spectrum of this model is
\begin{equation}
P_j=\frac{1- \alpha^2}{1 + \alpha^2 - 2 \alpha \cos(2 \pi \delta t / \tau_j)},
\label{eq18}
\end{equation}
where $\tau_j$ is the associated Fourier period (Eq.[\ref{eq16}]) of a scale $l_j$.  If the time series of interest is an AR(1) process, then a slice in the modulus of the continuous coefficients, $|\tilde f(l,t')|^2$, along dilation at a constant translation (the {\it local wavelet power spectrum}) should have have the form in Equation (\ref{eq18}).  Note that a white noise process is given by $\alpha=0$.

Using the theoretical background spectrum given above, we can develop significance tests for the continuous transform.  The background spectrum is the mean power spectrum expected for the assumed noise process (i.e., the `background' process), against which we wish to compare the actual signal.  Physical processes that are the result of this background process will produce power spectra that are normally distributed about this mean background spectrum.  Assuming that the values in our time series $f(t)$ are normally distributed, and because the square of a normally distributed variable is chi-square distributed with one degree of freedom, we expect the wavelet power $|\tilde f|^2$ to be chi-square distributed with two degrees of freedom.  The additional degree of freedom comes from the fact that the wavelet coefficients $\tilde f$ are complex for the Morlet wavelet.  In addition, the expectation value for the wavelet power of a white noise time series is just the variance $\sigma^2$; this expectation value provides a convenient normalization.  Using this normalization, the distribution for the local wavelet power spectrum is
\begin{equation}
\frac{ \left| \tilde f(l_j,t'_i) \right|^2}{\sigma^2} \Rightarrow P_j \frac{\chi^{2}_{\nu}}{\nu},
\label{eq19}
\end{equation}
where `$\Rightarrow$' means `is distributed as' and $\nu$ is the degrees of freedom, in this case two.  The indices on the scale $l$ run from $j=1,2, \ldots ,J$, where $J$ is the number of scales, and the indices on the translation $t'$ run from $i=1,2,\ldots ,N_{data}$.

We can also define the time-averaged wavelet power spectrum, or the {\it global wavelet spectrum}, as
\begin{equation}
\tilde f^{2}_{G} (l_j) = \frac{1}{N_j} \sum_{i=i_j}^{i'_j} \left| \tilde f (l_j, t'_i) \right|^2,
\label{eq20}
\end{equation}
where $i_j$ and $i'_j$ are the indices of the initial and final translations $t'_i$ outside of the cone of influence at a given scale $l_j$, and $N_j$ is the number $t'_i$ outside the cone of influence at that scale.  It has been shown that the global wavelet spectrum provides an efficient estimation of the true power of a time series \citep{per95}.  Averaging the wavelet power spectrum as in Equation (\ref{eq20}) increases the significance of the peaks, as the degrees of freedom is increased beyond what is used in the local wavelet power spectrum.  However, because the coefficients are correlated in both time and scale, the degrees of freedom for the global wavelet spectrum are
\begin{equation}
\nu_j = 2 \sqrt{1 + \left( \frac{N_j \delta t}{l_j \delta t_0} \right)^2},
\label{eq21}
\end{equation}
where $\delta t_0$ describes the decorrelation length in time.  For the Morlet wavelet of $k=6$, $\delta t_0 = 2.32$.  The global wavelet spectrum of a sinusoid is shown in Figure \ref{fig2}.

\setcounter{figure}{1}
\begin{figure}
\begin{center}
\scalebox{0.75}{\rotatebox{90}{\plotone{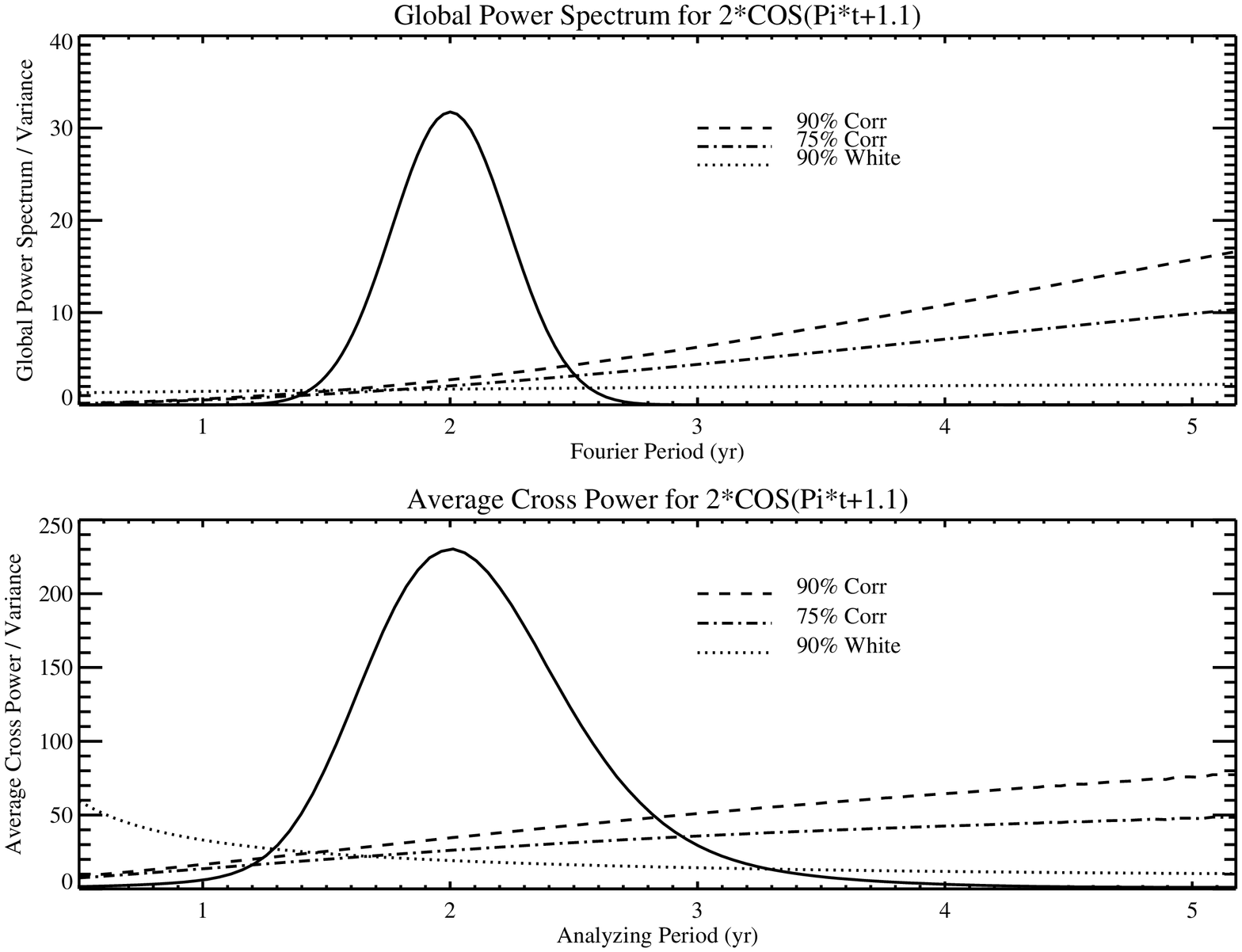}}}
\figcaption{Global wavelet power spectrum $\tilde f^{2}_{AG} (l_j)$ and the average cross power $\bar{F_c} (\tau_n)$ for a sinusoid of period $\tau_a = 2.0$ yr.  Also shown are the confidence levels for white and correlated noise, assuming a lag-1 autocorrelation of $\alpha = 0.911362$.\label{fig2}}
\end{center}
\end{figure}

Alternatively, we can smooth the wavelet spectrum in scale.  We define the scale-averaged wavelet power as
\begin{equation}
\tilde f^{2}_{L} (t'_i) = \sum_{j=j_1}^{j_2} \frac{\left| \tilde f (l_j, t'_i) \right|^2}{l_j}.
\label{eq22}
\end{equation}
The scale-averaged wavelet power can be interpreted as a time series of the average variance in the band $l_{j_1} \leq l_j \leq l_{j_2}$.  This distribution can be modeled as
\begin{equation}
\frac{\tilde f^{2}_{L} (t'_i)}{\sigma^2} \Rightarrow \bar{P} \frac{\chi^{2}_{\nu_l}}{\nu_l},
\label{eq23}
\end{equation}
where the scale-averaged theoretical spectrum $\bar{P}$ is
\begin{equation}
\bar{P} = \sum_{j=j_1}^{j_2} \frac{P_j}{l_j}.
\label{eq24}
\end{equation}
The degrees of freedom $\nu_l$ in Equation (\ref{eq24}) are modeled as
\begin{equation}
\nu_l = \frac{2 N_l L_{avg}}{L_{mid}} \sqrt{1 + \left( \frac{N_l \delta j}{\delta j_0} \right)^2},
\label{eq25}
\end{equation}
where
\begin{eqnarray*}
L_{avg} & = & \left( \sum_{j=j_1}^{j_2} \frac{1}{l_j} \right)^{-1} \\
L_{mid} & = & l_{min} 2^{0.5(j_1+j_2) \delta j} \\
l_j & = & l_{min} 2^{(j-1) \delta j}, \ \ j=1,2,\ldots, J, \\
\end{eqnarray*}
$N_l$ is the number of dilations averaged over, and $\delta j_0$ is the decorrelation distance in scale.  For the Morlet wavelet of $k=6$, $\delta j_0 = 0.60$.  The $\delta j$ describes how the dilations $l_j$ are constructed; it is common to construct them as fractional powers of two as it allows more emphasis on the smaller scales.  The factor $L_{avg}/L_{mid}$ corrects for the loss of degrees of freedom from dividing the wavelet power spectrum by scale in Equation (\ref{eq22}).

\subsection{Tests for the Cross Transform}

Using the results of \citet{tor98} from the previous section, we can derive significance tests for the cross-wavelet transform.  The power of the cross-wavelet $\tilde f_c$ can be written as
\begin{equation}
\left| \tilde f_c (l_j,t'_i) \right|^2 = \left| \tilde f_a (l_j, t'_i) \tilde f^{*}_{m} (l_j, t'_i) \right|^2 = \left| \tilde f_a (l_j, t'_i) \right|^2 \left| \tilde f_m (l_j, t'_i) \right|^2,
\label{eq26}
\end{equation}
and is distributed as
\begin{equation}
\frac{\left| \tilde f_a (l_j, t'_i) \right|^2 \left| \tilde f_m (l_j, t'_i) \right|^2}{\sigma^2} \Rightarrow P_j \left| \tilde f_m (l_j, t'_i) \right|^2 \frac{\chi^{2}_{\nu}}{\nu},
\label{eq27}
\end{equation}
where the degrees of freedom $\nu$ is two.  We can also define the cross-wavelet global power spectrum,
\begin{equation}
\tilde f^{2}_{CG} (l_j) = \frac{1}{N_j} \sum_{i=i_j}^{i'_j} \left| \tilde f_c (l_j, t'_i) \right|^2.
\label{eq28}
\end{equation}
Noting that the power of the continuous coefficients for the type of mock signal given in Equation (\ref{eq07}) is independent of the translation $t'$, the time-averaged form of Equation (\ref{eq13}), $F_c$,  then becomes
\begin{equation}
\bar{F_c} (\tau_m)= \sum_{j=1}^{J} \frac{\tilde f^{2}_{CG} (l_j)}{l_j} = \sum_{j=1}^{J} \frac{\tilde f^{2}_{AG} (l_j) \left|\tilde f_m (l_j) \right|^2}{l_{j}},
\label{eq29}
\end{equation}
where $\tilde f^{2}_{AG} (l_j)$ is the global wavelet power spectrum for the actual signal $f_a (t_i)$.  This distribution can be modeled as
\begin{equation}
\bar{F_c}(\tau_m) \Rightarrow \sum_{j=1}^{J} \frac{P_j \left|\tilde f_m (l_j) \right|^2}{l_{j}} \left(\frac{\chi^{2}_{\nu'_j}}{\nu'_j} \right),
\label{eq30}
\end{equation}
where the degrees of freedom $\nu'_j$ are modeled as
\begin{equation}
\nu'_j = \nu_j + \nu_l,
\label{eq31}
\end{equation}
for $\nu_j$ and $\nu_l$ given in the previous section.  This allows us to smooth the wavelet power in both time and scale, thus increasing the degrees of freedom, and to receive information of the characteristic time scale(s) of our time series from the mock coefficients $\tilde f_m$ rather than from the global wavelet spectrum.  As mentioned in \S~3.3, Equation (\ref{eq29}) allows us to interpret the cross-wavelet of a time series and an analyzing signal as a filter in wavelet space, where the coefficients of the continuous transform for the analyzing signal filter the global wavelet spectrum of our time series, $\tilde f^{2}_{AG}$.  Averaging the cross transform in scale then allows us to find the analyzing signal with which the time series is best correlated over the entire range of time and scale.  A plot of $\bar{F_c}$ for a sinusoid can be seen in Figure \ref{fig2}.

The results of this section are only valid when the cross-wavelet consists of one source that has an assumed background spectrum that is chi-square distributed.  If one is crossing two sources, $f_1 (t)$ and $f_2 (t)$, that have assumed background spectra, $P_1$ and $P_2$, that are chi-square distributed, then the cross-wavelet is distributed as
\begin{equation}
\frac{\left| \tilde f_1 (l,t) \tilde f^{*}_{2} (l,t) \right|}{\sigma_1 \sigma_2} \Rightarrow \frac{Z_{\nu} (p)}{\nu} \sqrt{P_{1} P_{2}},
\label{eq32}
\end{equation}
where $\sigma_1$ and $\sigma_2$ are the respective standard deviations, the degrees of freedom $\nu$ is two for complex wavelets, and $Z_{\nu} (p)$ is the confidence level for a given probability $p$ for the square root of the product of two chi-square distributions \citep{tor98}.  One can find the confidence level $Z_{\nu}$ by inverting the integral $p=\int_{0}^{Z_{\nu}} f_{\nu} (z) dz$.  The probability distribution is given by
\begin{equation}
f_{\nu}(z) = \frac{2^{2-\nu}}{\Gamma^2 (\nu / 2)} z^{\nu-1} K_0 (z),
\label{eq33}
\end{equation}
where $z$ is the random variable, $\Gamma$ is the Gamma function, and $K_0 (z)$ is the modified Bessel function of order zero.

\section{ALGORITHM FOR IMPLEMENTING THE CROSS TRANSFORM TO ANALYZE QUASIPERIODIC BEHAVIOR AND SIMULATIONS}
\subsection{Algorithm for Analysis of Quasiperiodic Behavior}

When using real data, we have no a priori knowledge, if any, of the periodicity for the
particular signal of interest, and require an algorithm that will allow
us to find it.  The results of \S~3.3 and \S~4 will serve as a guide
in developing this.  First, we subtract the mean from the signal and perform the continuous wavelet transform given by Equation (\ref{eq01}), only multiplying by a factor of $\sqrt{\delta t}$ to allow proper normalization and replacing the integral with a sum.  Then, we divide the continuous coefficients by the standard deviation of the time series to assure that the expectation value of $|\tilde f|^2$ for white noise is unity, and cross them with the coefficients for a number of
mock (analyzing) signals, each of which has a period $\tau_n = \tau_{n-1} + \delta \tau$.  For each of these, we average over the relevant region in wavelet
space as in Equation (\ref{eq29}), and look for extrema.  The extrema of $\bar{F_c} (\tau_n)$ then correspond to characteristic time scales, related by $\tau_a=0.973 \tilde \tau_n$, where $\tilde \tau_n$ is the period of the mock signal corresponding to the extremum.

The $n$th mock signal is given by
\begin{equation}
f_{n}\left(t\right)=A_n \cos\left(\frac{2 \pi}{\tau_{n}}t+\phi_{n}\right),
\label{eq34}
\end{equation}
where we use the subscript $n$ instead of $m$ on the parameters to emphasize that they are for the $n^{th}$ mock signal.  We choose a value of $A_n$ such that the global wavelet spectrum of the analyzing signal is normalized so that its maximum value is unity.  This provides a convenient normalization among the analyzing signals, assuring that they all have the same power.  In this analysis, 100 mock signals are used to analyze the actual time series, resulting in a step size of $\delta \tau \sim 0.045$ for most sources (this varies slightly because not all of the sources have the same time window).  We set $\phi_n$ equal to zero, as $|\tilde f_c|^2$ is independent of the phase difference between the signals.  Many of the signals analyzed in this paper have typical values for the time range on the order of $a=80$ yr and $b=101$ yr, which arise from using the year 1900 as a baseline.  The dilation values are chosen such that $2 \delta t < l < N_{data}\delta t/3 $ yr, which allows us to be
conservative when omitting edge effects and to admit the approximations used in
Equations (\ref{eq08}) and (\ref{eq13}).

When plotting the coefficients, those that fell within the cone of influence at any given dilation were set to a constant value, thus masking the edge effects.  Values of $l_j$ are shown logarithmically, so as to allow more sensitivity in the dilation region corresponding to shorter mock periods.  It is helpful to examine the global wavelet spectrum of the time series (Eq.[\ref{eq20}]), and to compare with the cross-wavelet results, so we include it in our analysis.  We assume a background spectrum $P_j$ of the form AR(1) (Eq.[\ref{eq18}]), and estimate the lag-1 autocorrelation coefficient by calculating the lag-1 and lag-2 autocorrelations, $\alpha_1$ and $\alpha_2$.  The lag-1 autocorrelation coefficient is then estimated as $\alpha = (\alpha_1 + \sqrt{\alpha_2})/2$.  The background spectrum $P_j$ then allows us to compute confidence levels using the results of \S~4.  We plot contours on the continuous and cross-wavelet transforms corresponding to the 90\% confidence level for an AR(1) process.  We also plot 90\% and 75\% correlated noise confidence levels on the plots of the global wavelet spectrum $f^{2}_{AG}$ and the averaged cross-wavelet power $\bar{F_c}(\tau_n)$, as well as 90\% confidence levels for white noise ($P_j=1$).  It should be noted that by confidence level, we mean that this given percentage of assumed background (i.e., noise) processes will produce behavior less than what is seen in the wavelet transforms.  For the time series analyzed here, the 90\% confidence levels for correlated noise mark where 90\% of AR(1) processes will produce values {\em less} than the 90\% values, and likewise for the 75\% confidence levels and the 90\% confidence levels for white noise.

To satisfy the admissibility condition (\ref{eq04}) we choose a value of $k=6$.  In reality, Equation (\ref{eq04}) is not formally satisfied for the Morlet Wavelet of Equation (\ref{eq01}).  However if $k \geq 5$, the admissibility condition is satisfied to within the accuracy of computer algorithms using single precision arithmetic.

\subsection{Simulations}

To test our technique, we ran several simulations.  First, to investigate the ability of our technique to detect a purely periodic signal in the presence of noise we used a signal of the form
\begin{equation}
f_{a}(t)=2 \sin(\pi t + 1.1)+Nn(t),
\label{eq35}
\end{equation}
where $N$ is an amplitude factor and $n(t)$ a noise
function that generates white noise.
We show the continuous and cross transforms for $N=10$ in Figure \ref{fig3}.  The cross wavelet plot is that for a mock signal of period $\tau_n =  2.059$ yr.  Plots of the global wavelet spectrum and $\bar{F_c}$ for this simulation are shown in Figure \ref{fig5}a.  As can be seen in Figure \ref{fig5}a, the period is easily recovered using the technique with noise of amplitude five-times that of the amplitude of the sinusoid (signal-to-noise of 0.2).  Because signal-to-noise values as low as $S / N = 0.2$ are never realized, our technique is is valid for the time series studied here.

\setcounter{figure}{4}
\begin{figure}
\begin{center}
\scalebox{1.2}{\rotatebox{90}{\plottwo{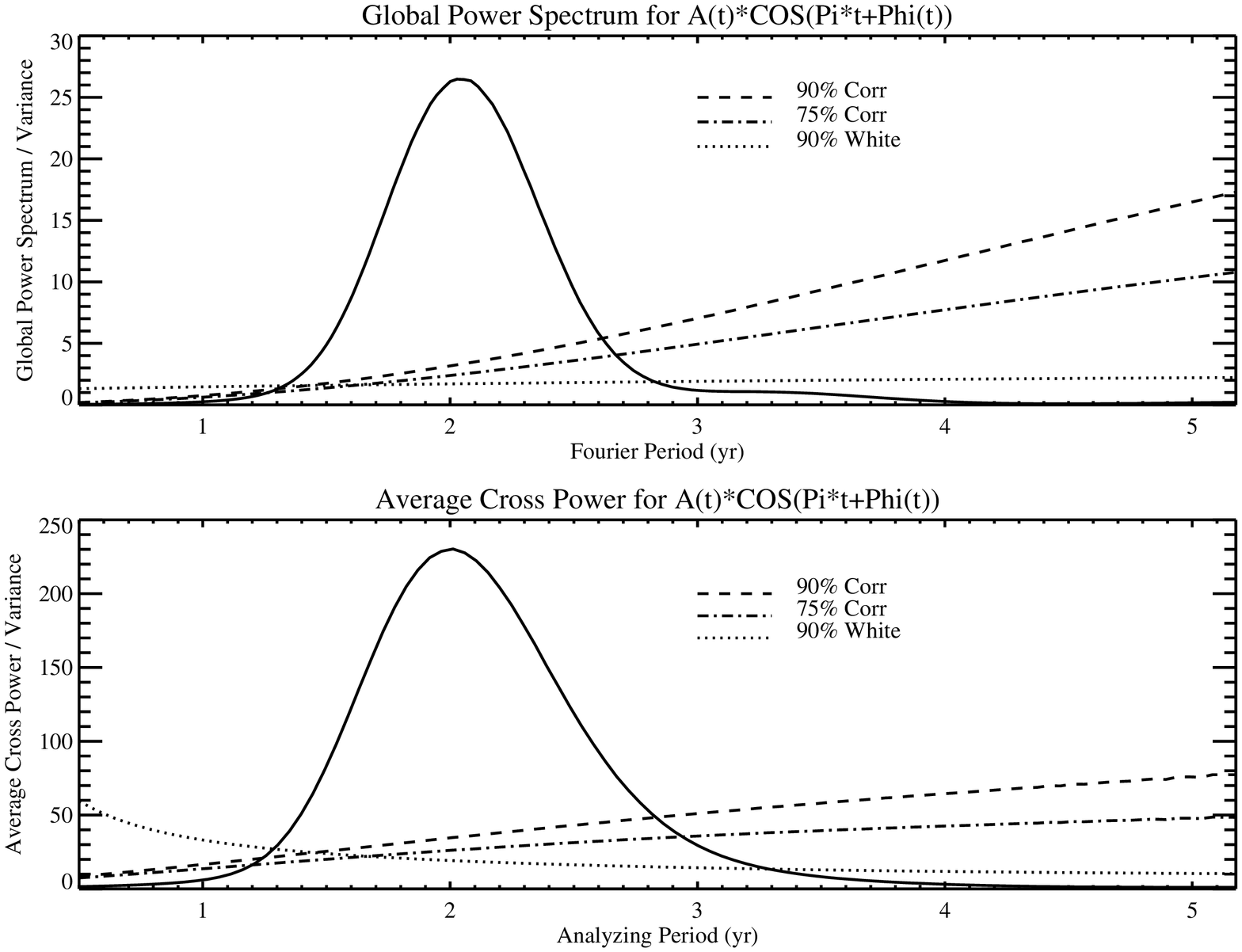}{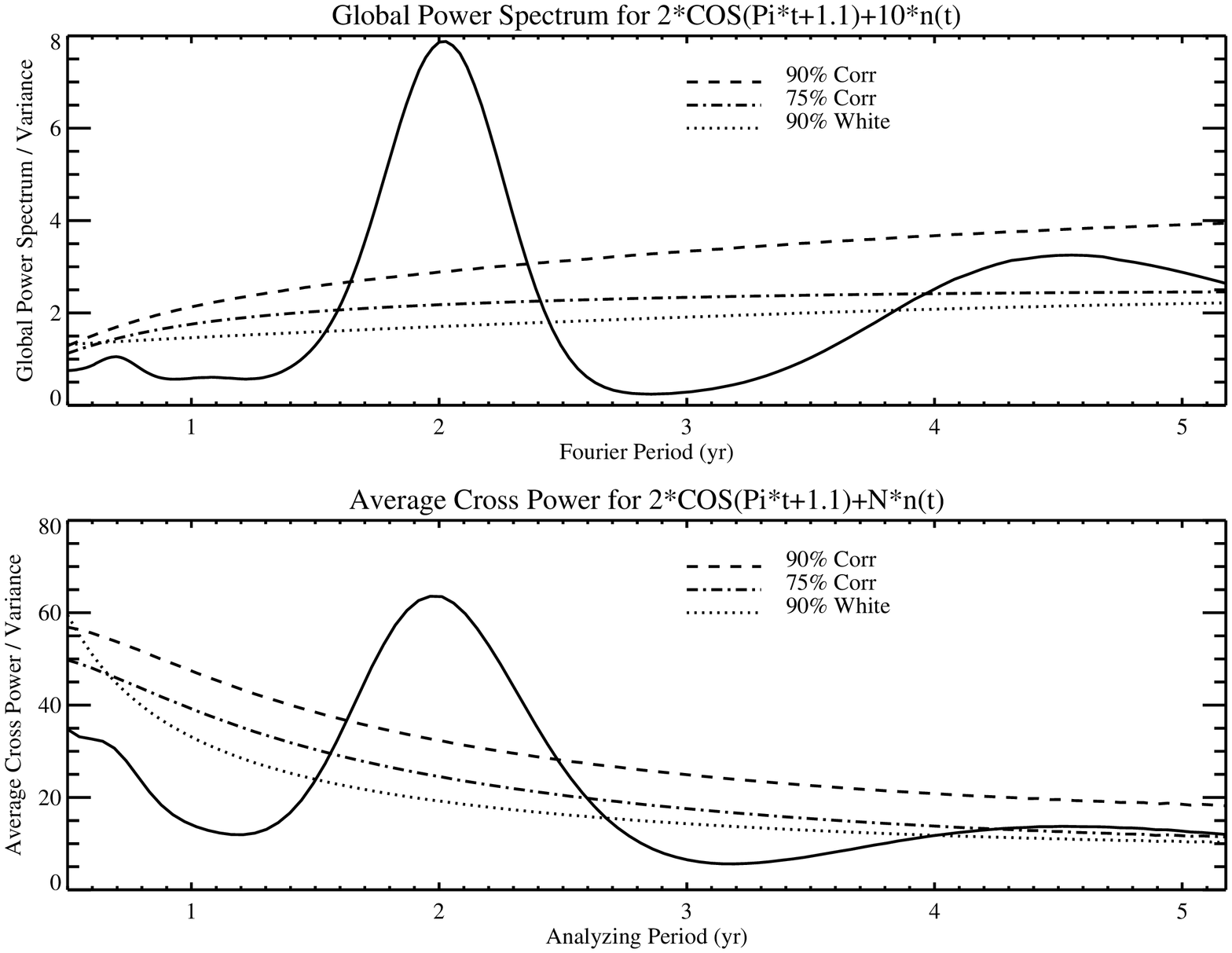}}}
\figcaption{Global power spectrum $\tilde f^{2}_{AG} (l_j)$ and average cross power $\bar{F_c} (\tau_n)$ for the sinusoids of Figure \ref{fig3} (top two panels) and  Figure \ref{fig4} (bottom two panels).  Confidence levels for correlated and white noise are shown.\label{fig5}}
\end{center}
\end{figure}

In addition, we also applied the algorithm to a sinusoid that changes amplitude and phase over the time window.  The transforms are shown in Figure \ref{fig4}, and the power spectra in Figure \ref{fig5}b.  A change in amplitude affects the amplitude of the wavelet coefficients, as would be expected.  The effects of a discontinuous change in the phase $\phi_a$ can be seen as disturbing the structure of the transforms in these regions.  The technique gives a characteristic period of 2.0 yr, showing that a sudden change in the phase of a signal does not effect the technique.

In practice, we are using this technique to investigate quasiperiodic behavior and to give a characteristic time scale for a time series, rather than attempting to recover a periodic signal buried beneath noise (although this certainly may be done).  To illustrate our technique in the case of pure noise, we applied the cross-wavelet technique to Gaussian white noise and correlated noise.  The wavelet transforms for a white noise signal are shown in Figure \ref{fig6}.  As can be seen, temporary quasiperiodic structure can arise for white noise, however the signal is clearly distinguished from more coherent signals by the complex structure throughout the plot, particularly in the low dilation (high frequency) region.  Plots of the global wavelet power spectrum and the averaged cross power $\bar{F_c}$ are shown in Figure \ref{fig8}a.  Although there are distinct peaks in the global spectrum at small scales, the average power of the cross-wavelet, $\bar{F_c}$, does not show distinct peaks at these scales.  We interpret this as being the result of smoothing in scale as well as time when calculating $\bar{F_c}$; the global wavelet spectrum is averaged in time for a given dilation $l_j$, and can result in peaks even in the case of non-stationary power, as is seen in the continuous plot for white noise.  However, by crossing with an analyzing signal, and averaging in dilation, we can smooth out spurious extrema, as is seen in the plot of $\bar{F_c}$.

\setcounter{figure}{7}
\begin{figure}
\begin{center}
\scalebox{1.2}{\rotatebox{90}{\plottwo{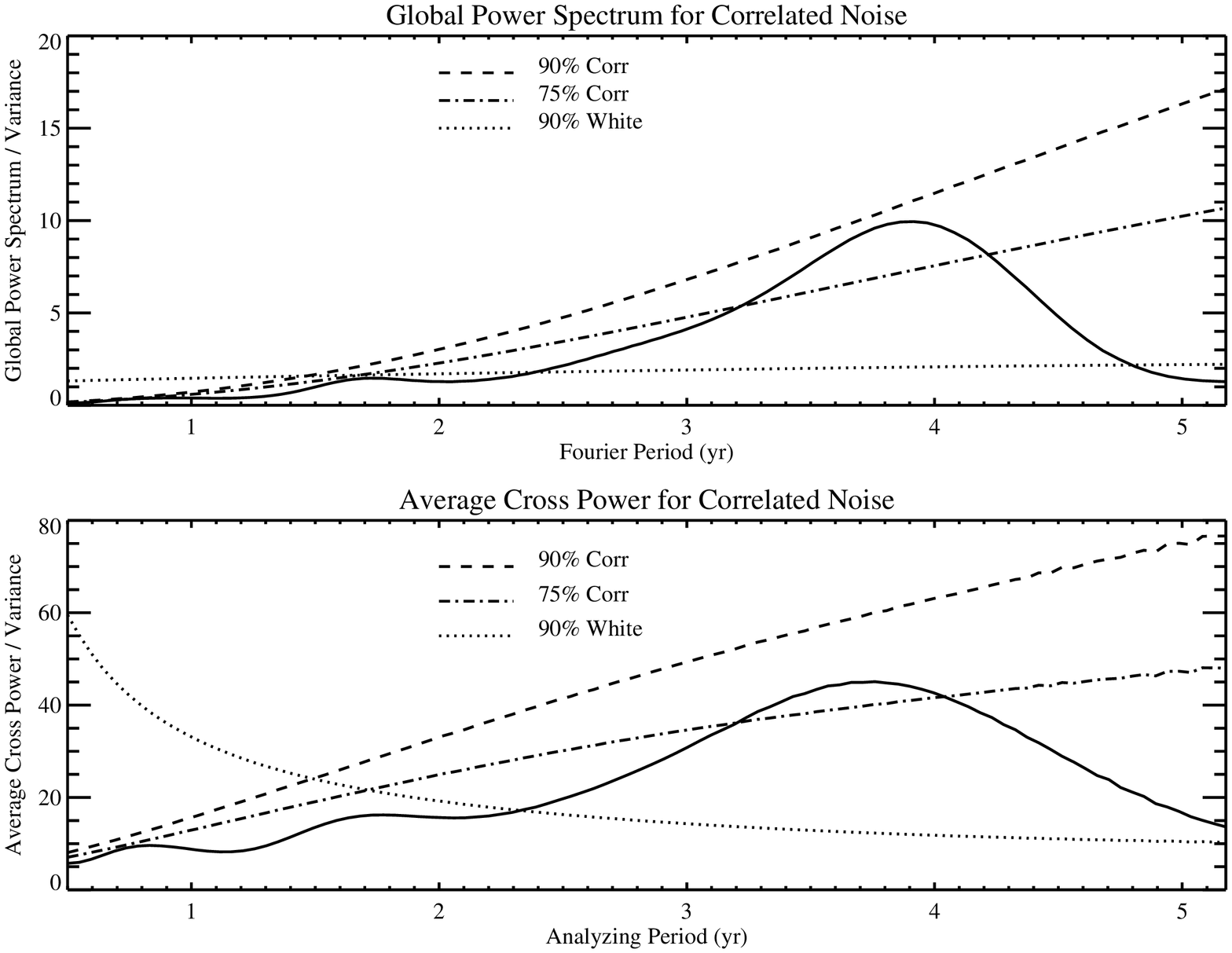}{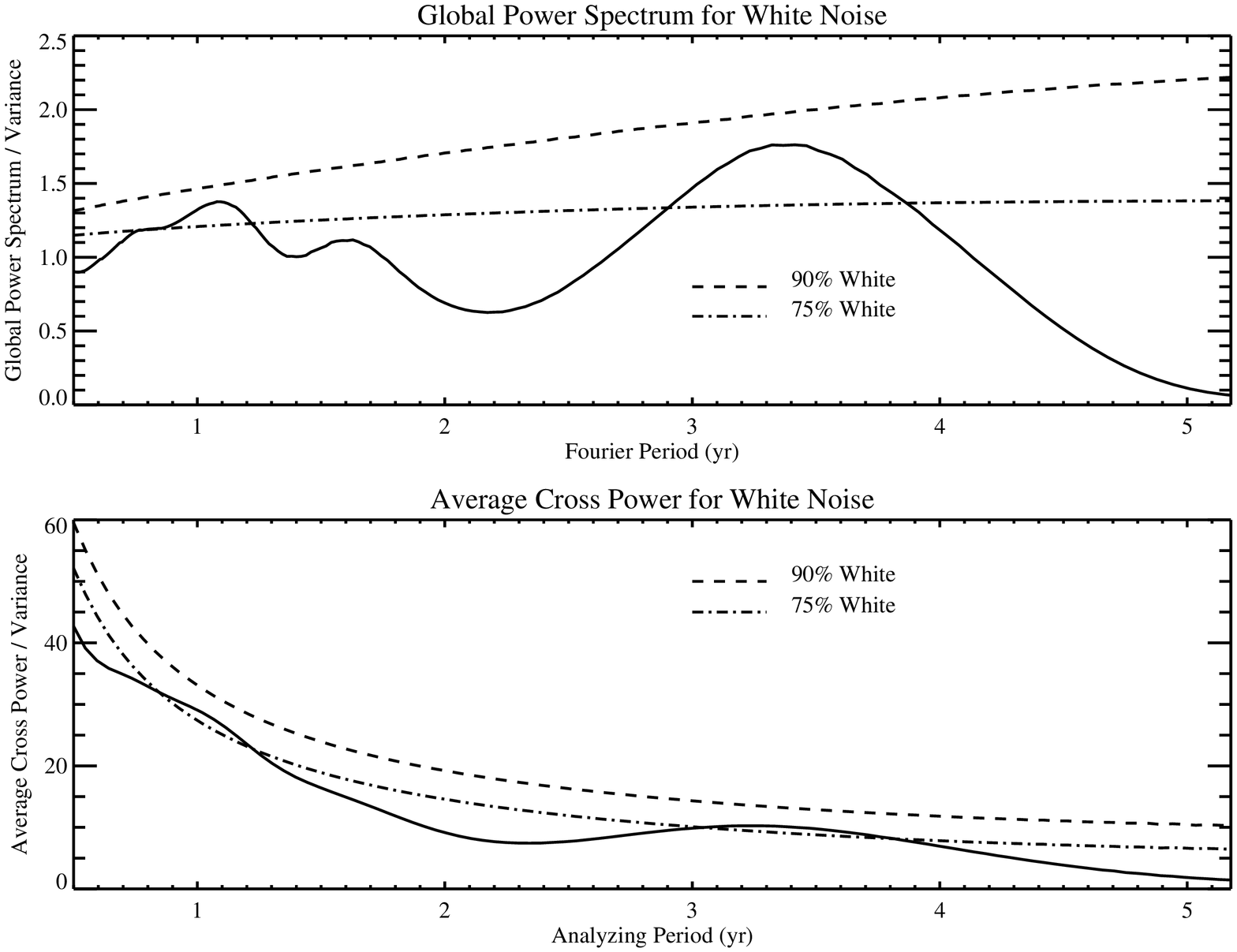}}}
\figcaption{Global power spectrum $\tilde f^{2}_{AG} (l_j)$ and average cross power $\bar{F_c} (\tau_n)$ for the white noise signal of Figure \ref{fig6} (top two panels) and the correlated noise signal of Figure \ref{fig7} (bottom two panels).  Confidence levels for correlated and white noise are shown.\label{fig8}}
\end{center}
\end{figure}

To illustrate the technique for correlated noise, we used a test signal of the type AR(1) with lag-1 autocorrelation $\alpha = 0.9$.  Many of the PR sources had lag-1 autocorrelations of $\alpha \sim 0.9$, and we found it helpful to simulate an AR(1) process of $\alpha$ similar to what we infer for the data.  Figure \ref{fig7} shows the transforms for this signal.  As expected, there is more activity at larger scales (lower frequencies).  Coherent activity does appear in the case of this type of correlated noise, however the power is not stationary and can not be characterized by a particular scale.  The global power spectrum and $\bar{F_c}$ can be seen in Figure \ref{fig8}b.  Peaks appear at similar periods for both, however the plot of $\bar{F_c}$ shows that the peaks are less significant than those seen in the global power spectrum.  The reason for this can be seen in the continuous plot; the power appears to split into two time scales around $t'=90$.  When we cross the continuous transform with our analyzing signal transform, and average over both translation and dilation, we see that the time scale of $\sim 3.8$ yr is not as significant as it appears in the global wavelet spectrum.

We can see from the simulations, as well as from the results of applying the transforms to the PR sources, that the average cross power, $\bar{F_c} (\tau_n)$, places more emphasis on the smaller scales and periods than the global power spectrum.  This is to be expected, as the amplitude of the global power spectrum for the time series of interest, $\tilde f^{2}_{AG} (l_j)$, has a dependence on the scale $l_j$ (see Eq.[\ref{eq08}]).  If we do not normalize the global power spectrum of the mock signal, $|\tilde f_m |^2$, such that its maximum value is unity for all periods $\tau_n$, then the amplitude of $|\tilde f_m |^2$ will also have a scale dependence.  This would result in the amplitude of $\bar{F_c}$ being dependent on the scale, $l_j$ as well (note that $\bar{F_c}$ would not be dependent on $l^2_j$ because we divide by the scale in Eq.[\ref{eq29}]).  However, normalizing the mock power spectrum $|\tilde f_m |^2$ so that it its maximum value is unity removes the scale dependence in $|\tilde f_m |^2$, and thus in $\bar{F_c}$.

In addition, we applied the cross-wavelet transform technique to the BL Lacs OJ 287 and AO 0235+164, as previous analysis has shown evidence for periodicity in these sources.  For OJ 287 the technique detected a characteristic time scale of $\sim 1.6$ yr, with a shorter time scale of $\sim 1.1$ yr, confirming results found earlier by \citet{hug98}.  The cross-wavelet analysis also confirmed earlier results for A0 0235+164, giving time scales of $\sim 1.9$ and $\sim 3.3$ yr, which is in good agreement with the $3.61$ yr found by \citet{roy00} using a Lomb-Scargle periodogram.  We were not able to confirm the longer periods found by \citet{roy00}, as they fell outside of our condition of at least four cycles in the time-window.

\section{THE DATA ANALYSIS}

The results of using the algorithm described in \S~5.1 to search UMRAO data on the Pearson-Readhead survey sources are given in Table \ref{tab1}.  Observations that we considered to have insufficient data to give a reliable analysis of quasiperiodicity (about half) are denoted by {\it D}.  Such observations were excluded because we required at least 100 data points for a given time series, as we assumed that 100 points are needed to adequately define its character.  When analyzing the light curves, we used the entire time window, which varied between observing frequencies and sources.  Since we are using a finite step size $\delta \tau \sim 0.045$ yr, we do not expect to use an analyzing signal with period $\tau_n=\tau_a$, but rather with period $\tau_n \approx \tau_a$.  However, because we are analyzing quasiperiodic behavior we do not expect the sources to have a well-defined period (i.e., a spike in Fourier space), but rather a characteristic time scale.  The error resulting from the finite step size of 0.045 yr is irrelevant in this context, as there is no sense found in assigning an extremely precise value for a quasiperiodic source.  We only record time scales at or above the 90\% confidence level.

\begin{figure}
\begin{center}
\scalebox{0.75}{\rotatebox{90}{\plotone{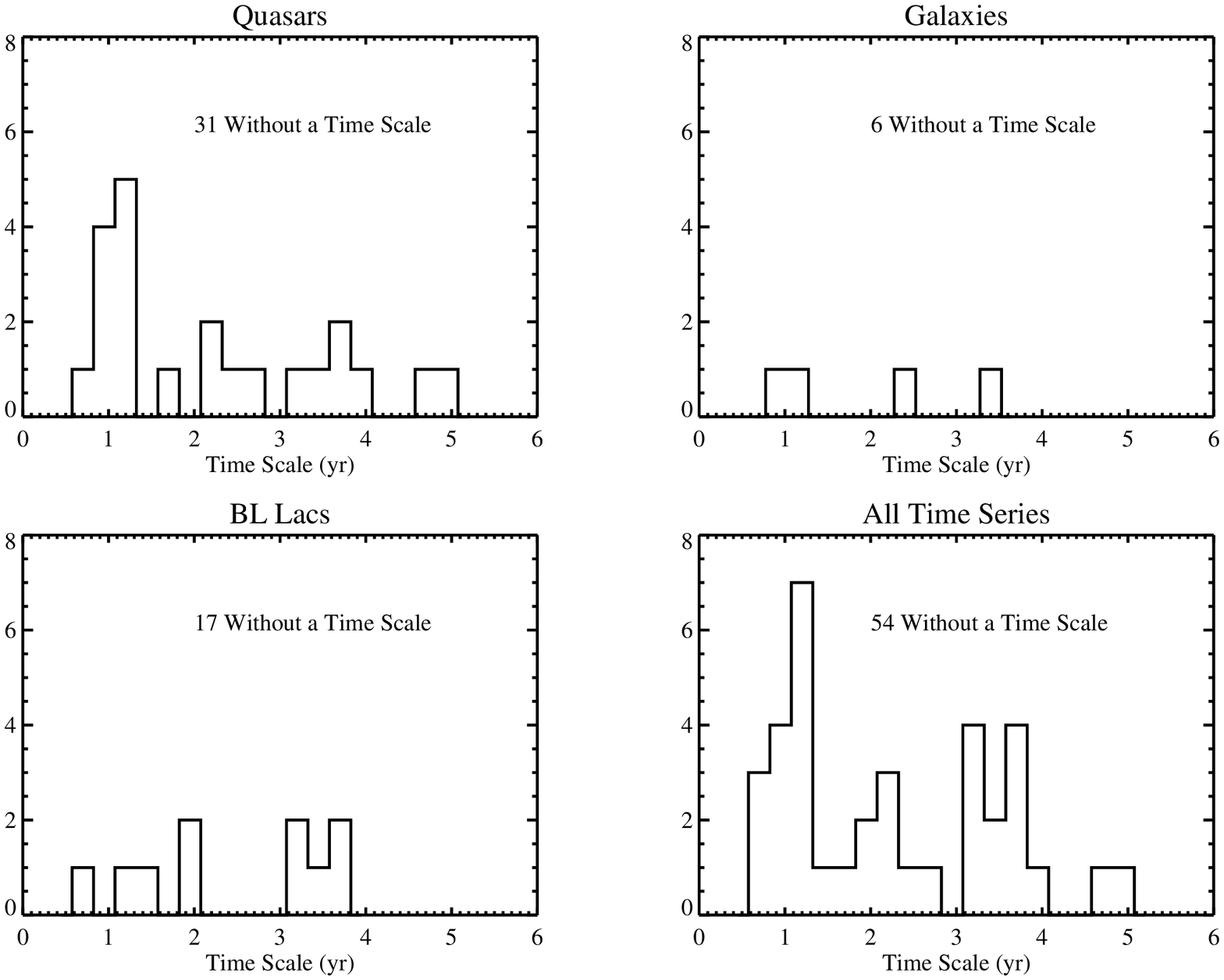}}}
\figcaption{Histograms for the quasiperiodicity statistics.  Time scales are binned every 0.25 years.\label{fig9}}
\end{center}
\end{figure}

Many of the sources that exhibited quasiperiodic variations had characteristic time scales between one and four years, with an average time scale for all quasiperiodic sources of 2.4 yr and a standard deviation of 1.3 yr.  This, of course, is to be expected as our observing interval is around twenty years, and we are only interested in variations with at least four possible repetitions across the interval.  Table \ref{tab2} shows statistics with respect to source type and Figure \ref{fig9} shows histograms for the number of characteristic periods found with respect to source type,  binned every 0.25 yr.  Out of thirty total sources with sufficient data, eighteen were quasars, four were galaxies, and eight were BL Lacs.  A little over half of the sources showed evidence for quasiperiodicity in at least one observing frequency.  Column six shows the ratio of recorded time scales to individual time series, which gives an idea of the average number of time scales per source.  For instance, if three quasars had sufficient data, and these quasars had eight individual time series between them over all the observing frequencies, with five of these individual time series exhibiting QPVs, then the ratio of recorded characteristic periods to individual time series would be $5 / 8$ for quasars.  A value of 0.33333 would mean that on average most of the sources for this particular type showed evidence for one characteristic time scale over the three observing frequencies, and two time scales over three observing frequencies for a value of 0.66667.  As can be seen from the table, on average there was about a little more than one characteristic period observed for every three observing frequencies for all sources.  The last column gives the ratio of sources with quasiperiodic behavior in more than one observing frequency to the total number of quasiperiodic sources.  Roughly half of the quasiperiodic sources exhibit quasiperiodicity across more than one observing interval.  It appears that quasars exhibit characteristic time scales more frequently than BL Lacs, as about two-thirds of the QSOs showed QPVs, while only about on-third of the BL Lacs did; however, all of the BL Lacs showing QPVs repeated the behavior in more than one observing frequency, while only half of the quasars did.  This may well reflect the fact that the BL Lacs of the PR survey sources generally have flat spectra \citep{all01}.  In addition, there does not appear to be a distinction among optical classes regarding the length of the time scales.  Among sources showing QPVs in more than one observing frequency, the characteristic time scales were similar among the different observing frequencies.  However, it is difficult to perform a complete statistical analysis of quasiperiodicity with respect to source type, as only 30 sources were analyzed.  In addition, there does not appear to be a distinct relationship between time scales and observing frequency; characteristic time scales do not consistently lengthen with increasing observing frequency or vice versa.  The plots of one particularly promising source for periodic behavior, the quasar 0804+499, can be seen in Figures \ref{fig10} and \ref{fig11} at an observing frequency of 4.8 GHz.

\setcounter{figure}{10}

\begin{figure}
\begin{center}
\scalebox{0.75}{\rotatebox{-90}{\plotone{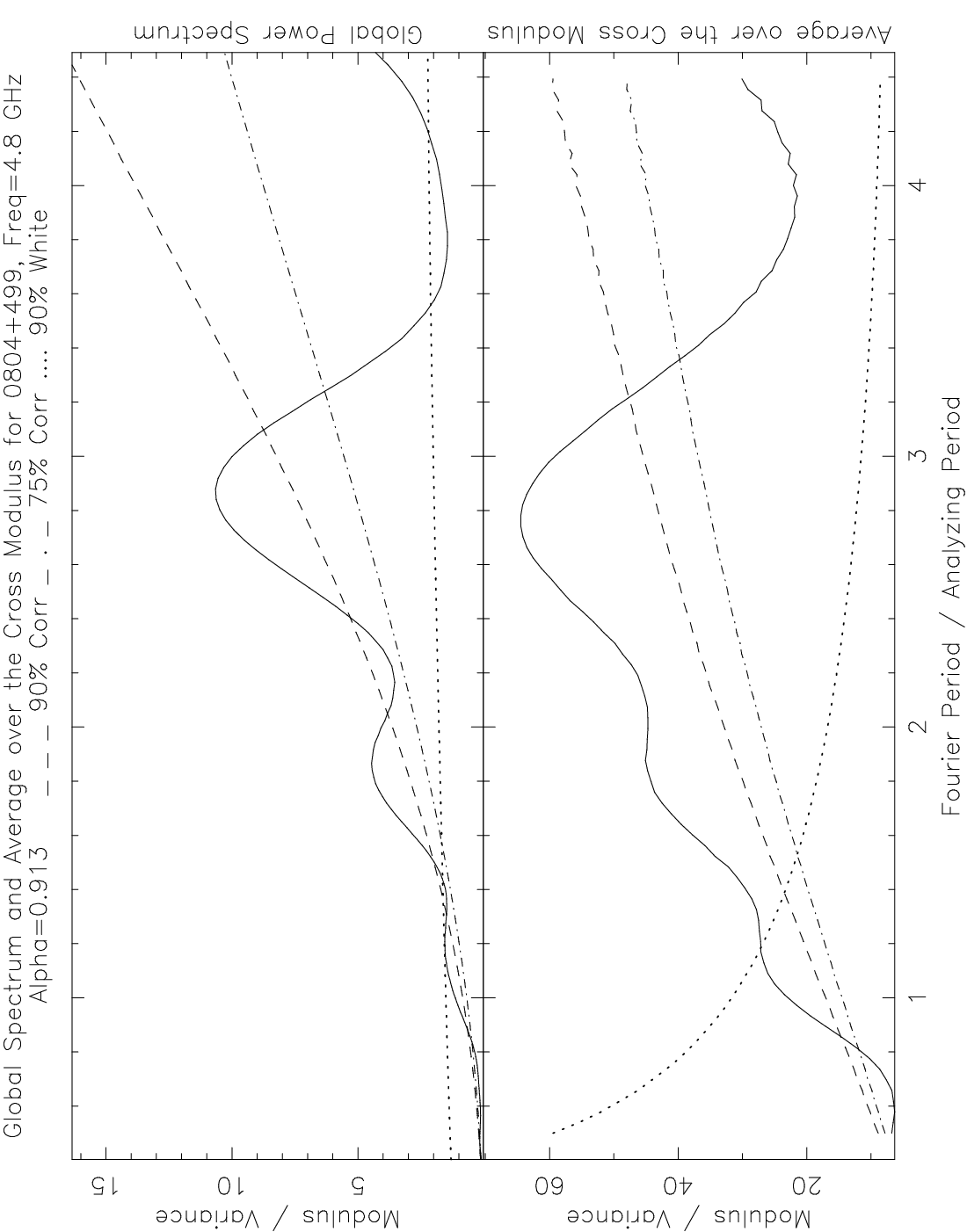}}}
\figcaption{Global power spectrum $\tilde f^{2}_{AG} (l_j)$ and the average cross power $\bar{F_c} (\tau_n) $ for 0804+499 at 4.8 GHz.  The three peaks corresponding to the three time scales seen in Figure \ref{fig10}a are seen in both plots.  The $\sim 2.7$ yr time scale is the most prevalent, as expected from visual inspection of the continuous plot.  Also shown are the confidence levels for correlated and white noise.\label{fig11}}
\end{center}
\end{figure}

Comparing the results for the PR sources with the results expected for correlated noise (\S~4), we find it unlikely that most of the PR sources with characteristic time scales are exhibiting behavior that can be attributed to an AR(1) process.  If the time series for the PR sources were the result of AR(1) processes, we would expect the ratio of recorded time scales to time series, $Q_t/T$, to be $\sim 0.1$, as we record characteristic time scales at or above the 90\% confidence level.  However, as can be seen in Table \ref{tab2}, $Q_t/T = 0.42$.

The lack of uniform quasiperiodic behavior across observing frequencies is unexpected for some of the sources with flat spectra.  We have explored the light curves for these sources and conclude that, while generally flat, they do exhibit behavior that is different across the three frequencies.  Localized activity is sufficient to remove or significantly weaken evidence for quasiperiodic behavior during that time interval, and if quasiperiodic behavior was noted at earlier or later times, it often did not span enough cycles (around four) to warrant recording a characteristic time scale.  Typically such events are not sufficient to affect the average spectral index, however they can result in changes in the small-scale structure of the light curves which is detected by the wavelet technique.  As an example, Figure \ref{fig12} shows the continuous transforms for 1823+568 at 8.0 GHz and 14.5 GHz, and the wavelet power spectra are shown in Figure \ref{fig13}.  Although the light curves appear similar, the continuous transforms show behavior that is different between the two observing frequencies.  Comparison with the power spectra and the continuous plots explains the discrepancy; the transform for the 8.0 GHz curve exhibits a transient time scale of $\sim 2.0$ yr, seen after $\sim$ 1993, whereas the 14.5 GHz curve exhibits a longer time scale of $\sim 3.3$ yr that persists throughout the time window.  We address the lack of common characteristic time scales with respect to observing frequency further in \S~7.

\setcounter{figure}{12}
\begin{figure}
\begin{center}
\scalebox{1.2}{\rotatebox{-90}{\plottwo{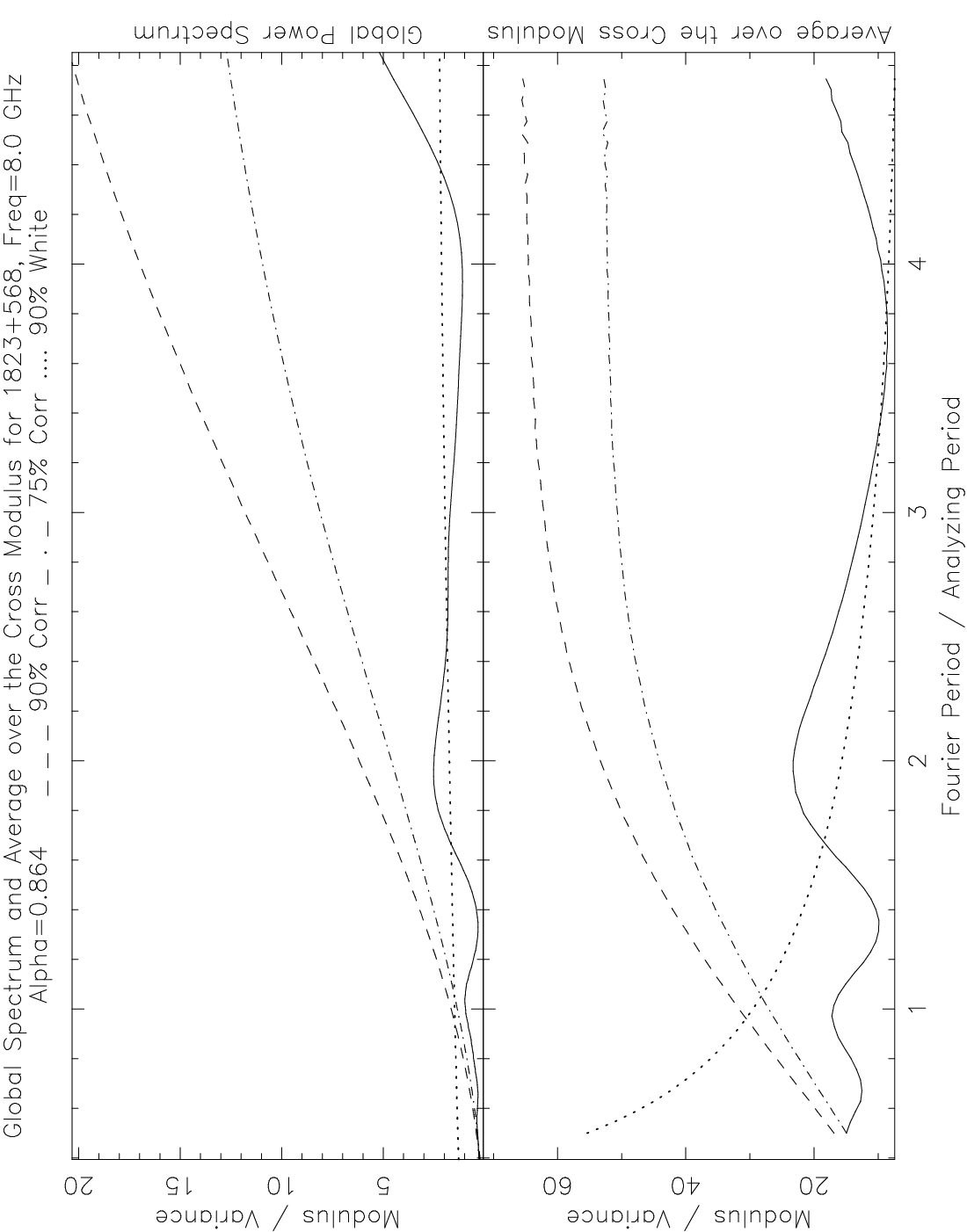}{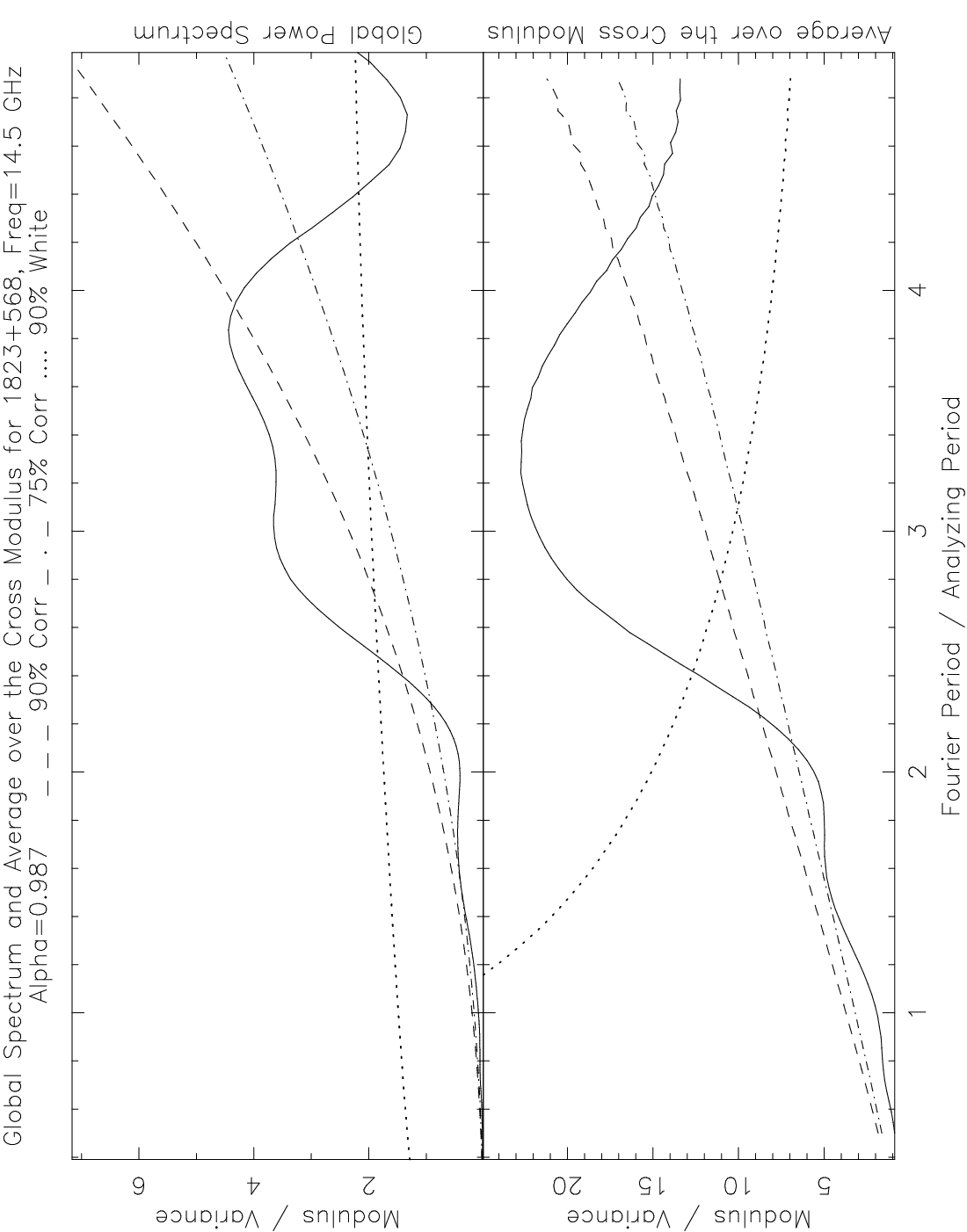}}}
\figcaption{Global power spectrum $\tilde f^{2}_{AG} (l_j)$ and the average cross power $\bar{F_c} (\tau_n)$ for 1823+568 at 8.0 GHz (top two panels) and 14.5 GHz (bottom two panels).  Although the $\sim 2.0$ yr time scale activity seen post-1990 in the continuous transform for the 8.0 GHz data is significant (see Figure \ref{fig12}a), these plots show that the source does not globally show any behavior in 8.0 GHz that is not inconsistent with an AR(1) process.  For the 14.5 GHz data, power is shared over a broad range in scale in both $\tilde f^{2}_{AG}$ and $\bar{F_c} (\tau_n)$, however the cross transform smooths the two peaks seen in the global spectrum into one distinct peak.  The single time scale interpretation given by the cross technique is visually reinforced in the continuous plot (see Figure \ref{fig12}b). \label{fig13}}
\end{center}
\end{figure}

In addition, we also note that for the 14.5 GHz data for 1823+568 the global wavelet spectrum $\tilde f^{2}_{AG} (l_j)$ and the average cross power $\bar{F_c} (\tau_n)$ differ in assigning a time scale.  The global spectrum reveals evidence for two time scales, one at $\sim 2.9$ yr and the other at $\sim 3.8$ yr, while the average cross power shows evidence for a time scale at $\sim 3.3$ yr.  Comparison between the two leads us to conclude that the characteristic time scale seen in the plot of $\bar{F_c}$ is more accurate.  The global wavelet spectrum shows a broad spread of power between the two time scales, where the time scales are inferred from the somewhat poorly resolved peaks.  The average cross power plot also shows a broad spread of power, but with only one peak.  This behavior is similar to what was described in \S~5.2 for the case of white noise; because the cross-wavelet technique smooths in dilation as well as in translation, and acquires information regarding time scales from the analyzing time series, it smooths the two poorly resolved peaks in the global wavelet spectrum into one broad peak.  Analyzing the continuous plot, we see that the plot of $\bar{F_c}$ better represents the behavior seen in the continuous transform, as power appears to shared among a range of scales of $\sim 1.0$ yr, rather than two distinct time scales separated by about a year.  It is more appropriate to assign one characteristic time scale in this case rather than two.  Behavior similar to what is seen in the 14.5 GHz data was seen in the 4.8 GHz data as well, but with less significance.

A comparison of the results from Table \ref{tab1}, with those from the structure function analysis performed by \citet{hug92} finds broad agreement, in the sense that most of the sources that we find to have characteristic time scales also exhibited time scales in the structure function analysis, and likewise for those sources lacking time scales.  After comparing the time scales from the structure function analysis for those sources that had time scales short enough to meet our requirement of four repetitions over the time window (i.e., shorter than 4-5 years for most sources), we find that many of the time scales we find are comparable to those found from the structure function analysis.  For instance, we deduce a time scale of $\sim 2.6$ yr for the source 0836+710 at 4.8 GHz, while the structure function finds that the time series for this source is not correlated above a time scale of $2.88$ yr; also, we find a characteristic time scale of $\sim 2.0$ yr for the BL Lac 1803+784, and the structure function analysis gives a time scale of $1.86$ yr.  Although the structure function analysis gives a measure of the time scale above which variations appear to be uncorrelated, which is not the same as the time scale that we measure, we find that the results found from the cross-wavelet technique agree with those found the earlier structure function analysis of \citet{hug92}.

Color postscript plots for the continuous and cross-wavelet transforms, as well as the plots for the global wavelet spectrum and the average cross power, are available for all sources from the UMRAO website at \\
{\bf http://www.astro.lsa.umich.edu/} \\
\ \ \ {\bf obs/radiotel/prcwdata.html}.

\section{CONCLUSIONS}

We conclude that, complimented with the continuous wavelet transform, the cross-wavelet technique can be an effective tool in the search for quasiperiodicity of a time series.  Comparing the results for the Pearson-Readhead survey sources with that expected for a correlated noise process of the form AR(1), we conclude that the observed quasiperiodic behavior is unlikely to be the result of an AR(1) process, as about 40\% of the time series for the PR sources had characteristic time scales compared to about 10\% expected for an AR(1) process.  The observed time scales may be the result of a type of correlated noise that is not AR(1), and it certainly may not even be stationary noise.  However, even if the QPVs arise from correlated noise, it is meaningful to explore the characteristic time scale of a time series, and such results provide a useful diagnostic of the underlying variations.

After applying the cross-wavelet algorithm to the Pearson-Readhead VLBI survey sources, analysis revealed evidence for quasi-periodic variations in $\sim 57\%$ of the sources, as well as evidence that $\sim 67\%$ of quasars, $\sim 38\%$ of the BL Lacs, and $50\%$ of the galaxies have quasiperiodic behavior in at least one observing frequency.  The sources were observed to have a mean characteristic time scale of 2.4 yr, with standard deviation of 1.3 yr.

Because the analyzed radio band variations for the PR sources originate in parsec-scale jets, perturbations that propagate at $c$ will lead to observable fluctuations if they span the flow, guaranteeing time scales of order years.  In addition, coherent perturbations that arise from the excitation of certain modes of oscillation of the flow, which could give rise to quasiperiodic behavior with time scales comparable to the dynamical response time of the flow (i.e., years), have been shown to arise quite naturally \citep{har01}.  We conclude that our results are in good agreement with the characteristic time scales that we would expect to observe based on the nature of these objects.

There is a transition region where the jet changes from optically thick to optically thin, with optical depth $\tau=1$ \citep{caw91}.  Because this region varies with the observing frequency, we are looking at a different physical location in the jet at each observing frequency.  It is likely that the time scale of quasiperiodic variations is dependent on their location in the jet.  Naively, we would expect to probe larger scale regions with longer
characteristic time scales, the lower the observing frequency chosen.  The
results of our analysis certainly find no such correlation of time scale and
frequency.  However, jets may accelerate due to adiabatic expansion, or
decelerate due to entrainment, leading to a change in bulk Lorentz factor
with position.  In addition, curvature is now known to be a common feature of these
flows. A change in flow speed and/or flow orientation with respect to the
observer can lead to a significant change in Doppler factor, and thus to the
observed time scale of activity, masking any simple trends. Furthermore,
local jet properties and ambient conditions play a major role in determining
what modes of the flow exist, and may be excited, and observations that probe
different physical scales might well reveal activity in one frequency band but
not in another.  

Although the number of sources analyzed here is in no way exhaustive, we see no reason why such quasiperiodic behavior should be confined to the Pearson-Readhead survey sources, and we find it likely that many active galactic nuclei exhibit quasiperiodicity.  Only four galaxies had sufficient data for analysis, and so are tabulated but not discussed.  A statistical analysis on the results of applying the cross-wavelet technique to a greater number of sources would lead to a more interesting and conclusive comparison of quasiperiodic behavior and source type, as well as evidence for quasiperiodicity in active galactic nuclei in general.  In addition, the cross-wavelet transform may also be a useful tool for analyzing the correlation of a source between different observing frequencies.

We would like to thank the anonymous referee for many helpful and insightful comments that have contributed to significant improvement of this manuscript.

\clearpage

\begin{deluxetable}{cccccccc}
\singlespace
\tablewidth{0pt}
\tabletypesize{\scriptsize}
\tablecaption{Characteristic periods found using the cross-wavelet algorithm on the Pearson-Readhead VLBI survey sources.  All periods are in years.\label{tab1}}
\tablehead{
\colhead{Source} & \colhead{Opt. Class} & \colhead{4.8 GHz} &
\colhead{Data} & \colhead{8.0 GHz} & \colhead{Data} &
\colhead{14.5 GHz} & \colhead{Data}}
\startdata 
0016+731 & Q & - & 218 & - & 170 & 3.3 & 249 \\ 
3C 20 & G & D & & D & & D & \\ 
0108+388 & G & D & & - & 125 & D & \\
DA 55 & Q & - & 566 & - & 811 & 2.3 & 674 \\ 
0153+744 & Q & D & & D & & D & \\ 
0212+735 & Q & - & 228 & - & 244 & 4.1 & 263 \\ 
3C 66B & G & D & & D & & D & \\ 
3C 83.1 & G & D & & D & & D & \\ 
3C 84 & G & - & 636 & - & 1617 & - & 872 \\ 
0404+768 & G & D & & D & & D & \\ 
3C 147 & Q & - & 141 & - & 173 & - & 134 \\ 
3C 153 & G & D & & D & & D & \\ 
OI 147 & G & D & & D & & D & \\ 
OI 318 & Q & - & 134 & 1.0 & 133 & 0.7 & 124 \\ 
3C 179 & Q & - & 136 & 1.0 & 125 & 2.2 & 134 \\ 
0804+499 & Q & 2.7, 1.8, 1.2 & 193 & 1.1 & 196 & 1.1 & 177 \\ 
3C 196 & Q & D & & D & & D & \\ 
0814+425 & BL & - & 177 & - & 462 & - & 282 \\ 
0831+557 & G & D & & D & & D & \\ 
0836+710 & Q & 2.6 & 222 & 1.4 & 166 & - & 271 \\ 
0850+581 & Q & D & & D & & D & \\
0859+470 & Q & D & & D & & D & \\ 
3C 216 & Q & - & 189 & - & 228 & - & 194 \\ 
3C 219 & G & D & & D & & D & \\ 
4C 39.25 & Q & - & 554 & - & 1282 & - & 815 \\ 
4C 20.24 & Q & D & & 3.6 & 141 & 3.7 & 104 \\ 
M 82 & G & D & & D & & D & \\ 
0954+556 & Q & D & & - & 124 & - & 103 \\ 
0954+658 & BL & - & 167 & - & 150 & - & 221 \\
3C 236 & G & D & & D & & D & \\ 
1031+567 & G & D & & D & & D & \\ 
3C 268.1 & G & D & & D & & D & \\  
3C 280 & G & D & & D & & D & \\
4C 66.22 & G & 1.3 & 168 & - & 176 & 0.9 & 147 \\ 
3C 295 & G & D & & D & & D & \\ 
3C 309.1 & Q & D & & D & & D & \\ 
3C 330 & G & D & & D & & D & \\ 
4C 41.43 & Q & D & & D & & D & \\ 
1633+382 & Q & - & 244 & - & 432 & - & 315 \\ 
3C 343 & Q & D & & D & & D & \\ 
1637+574 & Q & D & & D & & D & \\ 
3C 345 & Q & - & 851 & - & 1529 & 5.1 & 927 \\ 
1642+690 & Q & - & 167 & - & 331 & 4.9 & 278 \\
MKn 501 & BL & - & 230 & - & 525 & - & 302 \\ 
1739+522 & Q & 1.4 & 248 & 1.4 & 348 & 1.2 & 383 \\ 
1749+701 & BL & - & 217 & - & 357 & - & 256 \\ 
1803+784 & BL & 2.0, 3.9 & 272 & 2.0 & 300 & - & 419 \\
3C 371 & BL & - & 220 & - & 451 & - & 307 \\ 
1823+568 & BL & 3.3 & 135 & - & 241 & 3.3 & 275 \\ 
3C 380 & Q & D & & D & & D & \\ 
3C 388 & G & D & & D & & D & \\ 
3C 390.3 & G & D & & 3.4, 2.4 & 150 & - & 154\\
1928+738 & Q & - & 315 & - & 356 & - & 469 \\
3C 401 & G & D & & D & & D & \\
OV 591 & Q & D & & D & & D & \\ 
OW 673 & Q & D & & D & & D & \\ 
3C 438 & G & D & & D & & D & \\ 
BL LAC & BL & 1.4 & 856 & 3.7 & 1441 & 3.5, 1.6, 0.7 & 1243 \\ 
2229+391 & G & D & & D & & D & \\ 
3C 452 & G & D & & D & & D & \\ 
2351+456 & Q & - & 148 & D & & 3.8 & 167 \\ 
DA 611 & G & D & & D & & D & \\ 
\enddata
\tablecomments{D signifies insufficient data, - signifies no detected quasiperiodicity.  The number of data points are given only for those sources with sufficient data.  If two time scales are given, the more prominent one is listed first.}
\end{deluxetable}

\clearpage

\begin{deluxetable}{cccccccc}
\singlespace
\tablewidth{0pt}
\tabletypesize{\scriptsize}
\tablecaption{Quasiperiodicity Statistics. \label{tab2}}
\tablehead{
\colhead{Type}  & \colhead{Sources} & Time Series & \colhead{Mean Time Scale (yr)} &
\colhead{Stnd. Dev. (yr)} & \colhead{$Q/S$} & \colhead{$Q_t/T$} & \colhead{$M/Q$} \\ 
1 & 2 & 3 & 4 & 5 & 6 & 7 & 8}
\startdata 
Quasars & 18 & 51 & 2.3 & 1.4 & 0.67 & 0.43 & 0.5 \\
Galaxies & 4 & 10 & 2.0 & 1.1 & 0.5 & 0.4 & 0.5 \\
BL Lacs & 8 & 24 & 2.3 & 1.1 & 0.38 & 0.42 & 1.0 \\
All Types & 30 & 85 & 2.4 & 1.3 & 0.57 & 0.42 & 0.59
\enddata
\tablecomments{$Q/S$ is the ratio of sources exhibiting quasiperiodicity in at least one observing frequency to the total number of sources.  $Q_t/T$ is the ratio of the total number of time scales to the total number of time series.  $M/Q$ is the ratio of sources that showed quasiperiodicity in more than one observing frequency to the total number of quasiperiodic sources.}
\end{deluxetable}

\setcounter{figure}{0}

\begin{figure}
\begin{center}
\figcaption{(a) Continuous wavelet transform for a sinusoid of period $\tau_a = 2.0$ yr. (b) Cross-wavelet transform for the sinusoid with another sinusoid of period $\tau_m=2.059$ yr.  The contours on both power (modulus) plots are for the 90\% confidence level for correlated noise, assuming a lag-1 autocorrelation of $\alpha = 0.911362$.\label{fig1}}
\end{center}
\end{figure}
\setcounter{figure}{2}
\begin{figure}
\begin{center}
\figcaption{(a) Continuous wavelet transform for a sinusoid of period $\tau_a = 2.0$ yr with white noise added, $S/N=0.2$. (b) Cross-wavelet for the sinusoid of (a).  The contours on both power plots are for the 90\% confidence level for correlated noise, assuming a lag-1 autocorrelation of $\alpha = 0.283620$.\label{fig3}}
\end{center}
\end{figure}

\begin{figure}
\begin{center}
\figcaption{(a) Continuous wavelet transform for a sinusoid of period $\tau_a = 2.0$ yr and varying amplitude and phase. (b) Cross-wavelet transform for the sinusoid of (a).  The contours on both power plots are for the 90\% confidence level for correlated noise, assuming a lag-1 autocorrelation of $\alpha = 0.894581$.\label{fig4}}
\end{center}
\end{figure} 
\setcounter{figure}{5}
\begin{figure}
\begin{center}
\figcaption{(a) Continuous wavelet transform for a Gaussian white noise signal.  White noise is easily recognizable by its erratic and complex structure in wavelet space, particularly at small dilations.  (b) Cross-wavelet transform for the Gaussian white noise signal, corresponding to an analyzing signal of period $\tau_n=0.736$ yr.  The contours on both power plots are for the 90\% confidence level for white noise. \label{fig6}}
\end{center}
\end{figure}

\begin{figure}
\begin{center}

\figcaption{(a) Continuous wavelet transform for a correlated noise signal of the form AR(1), with lag-1 autocorrelation coefficient $\alpha = 0.9$. (b) Cross-wavelet transform for the correlated noise signal.  The signal is crossed with a mock signal of period $\tau_n = 3.901$ yr.  The contours on both power plots are for the 90\% confidence level for an AR(1) process of $\alpha=0.9$.\label{fig7}}
\end{center}
\end{figure}
\setcounter{figure}{9}
\begin{figure}
\begin{center}

\figcaption{(a) Continuous wavelet transform for 0804+499 at 4.8 GHz.  The periodic behavior is easily seen as repetition in the real plot.  The modulus plot shows the significance of the behavior; particularly notable is the $\sim 2.7$ yr period. (b) Cross-wavelet transform for 0804+499 at 4.8 GHz with an analyzing signal of period $\tau_n = 2.86$ yr.  The data is well correlated with a sinusoid of this period, as is evidenced by the uniformity in the cross-wavelet plots.  The contours on both power plots are for the 90\% confidence level for correlated noise, assuming a lag-1 autocorrelation of $\alpha = 0.913$.\label{fig10}}
\end{center}
\end{figure}
\setcounter{figure}{11}
\begin{figure}
\begin{center}

\figcaption{(a) Continuous wavelet transform for the BL Lac 1823+568 at 8.0 GHz.  The contours are the 90\% confidence level for an AR(1) process of $\alpha = 0.864$. Note the significant behavior seen after 1990. (b) Continuous wavelet transform for 1823+568 at 14.5 GHz.  The contours are the 90\% confidence level for an AR(1) process of $\alpha = 0.987$.  One can see that power is distributed over a broad range in scale for this plot, centered about a dilation corresponding to a time scale of $\sim 3.3$ yr.  Although the two time series share similarities, there are enough differences to account for the varying behavior seen in the transforms.\label{fig12}}
\end{center}
\end{figure}

\end{document}